\documentclass[journal=jacsat,manuscript=article]{achemso}

\usepackage[version=3]{mhchem} 
\usepackage{color}
\usepackage{enumerate}

\author{Lea M. Ibele}
\affiliation{Universit\'e Paris-Saclay, CNRS, Institut de Chimie Physique UMR8000, 91405, Orsay, France}
\altaffiliation{Aix Marseille University, CNRS, ICR, 13397 Marseille, France}
\email{lea-maria.ibele@univ-amu.fr}
\author{Eduarda Sangiogo Gil}
\affiliation{Institute of Theoretical Chemistry, Faculty of Chemistry, University of Vienna, W\"ahringer Str. 17, 1090 Vienna, Austria}
\email{eduarda.sangiogo.gil@univie.ac.at}
\author{Peter Sch\"urger}
\affiliation{Universit\'e Paris-Saclay, CNRS, Institut de Chimie Physique UMR8000, 91405, Orsay, France}
\author{Rodrigue Noc}
\affiliation{Universit\'e Paris-Saclay, CNRS, Institut de Chimie Physique UMR8000, 91405, Orsay, France}
\author{Federica Agostini}
\affiliation{Universit\'e Paris-Saclay, CNRS, Institut de Chimie Physique UMR8000, 91405, Orsay, France}
\email{federica.agostini@universite-paris-saclay.fr}

\SectionNumbersOn

\title{A coupled-trajectory strategy for decoherence, frustrated hops and internal consistency in surface hopping}

\abbreviations{IR,NMR,UV}

\begin{document}

\begin{tocentry}
\centering
\includegraphics[width=\textwidth]{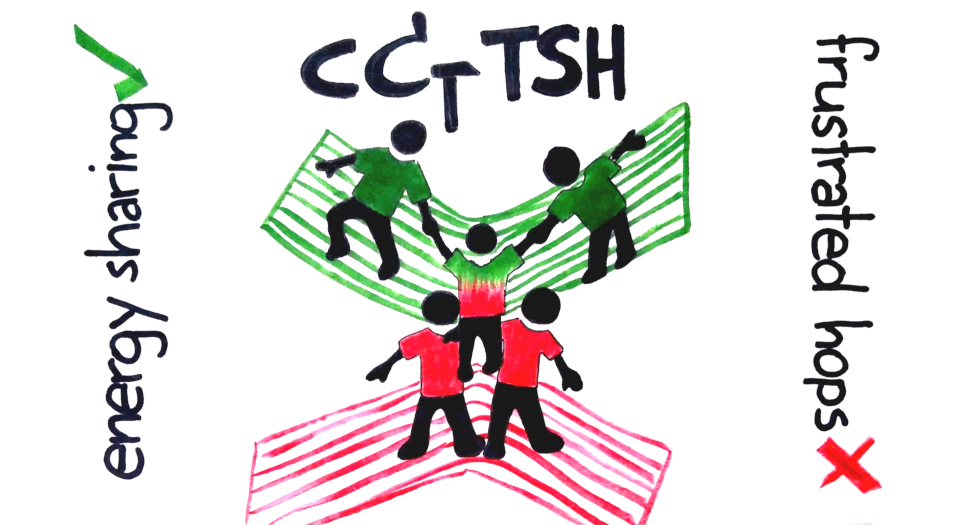}
\end{tocentry}

\begin{abstract}
We address the issues of decoherence, frustrated hops and internal consistency in surface hopping. We demonstrate that moving away from an independent-trajectory picture is the strategy which allows us to propose a robust surface hopping scheme overcoming all these issues at once. Based on the exact factorization and on the idea of coupled trajectories, we consider the swarm of trajectories, that mimics the nuclear dynamics in nonadiabatic processes, as a unique entity. In this way, imposing energy conservation of the swarm and allowing the trajectories to share energy when hops occur clearly indicates the route towards a new surface hopping scheme. Encouraging results are reported, in terms of electronic and vibrational time-dependent properties on the photodynamics of fulvene and  4-(dimethyloamino)benzonitrile, modeled with a full-dimensional linear vibronic coupling Hamiltonian.
\end{abstract}


\section{Introduction}
Nonadiabatic molecular dynamics is an irreplaceable method in the toolkit of physical chemists to model time-resolved photochemistry, the study of the coupled electron-nuclear evolution of molecules upon light excitation\cite{book_Gonzalez_Lindh}. Among the many theories available nowadays, surface hopping, since its first appearance in the literature in 1971\cite{tully1971trajectory}, has grown to become \textsl{the} method of choice to simulate the nonadiabatic dynamics of molecules and to accompany experimental observations.\cite{Tully_JCP1990, ibele2020excimer, pathak2020tracking, squibb2018acetylacetone, avagliano2022automatized, Prezhdo_JPCL2016, Akimov_CCC2022, Tully2012, Qiu2023} Despite alternative theories being continuously developed, surface hopping remains very popular thanks to the unrivaled accuracy-cost compromise, to its relative implementation simplicity, and especially to the fact that it conveys a very intuitive picture of a nonadiabatic process.\cite{Gonzalez_WIRES2018, barbatti2011nonadiabatic, agostini2019different,Barbatti_CR2018, PERSICO2024273, Barbatti_WIRES2011}
In surface hopping, the nuclear densities associated with the electronic states participating in the photochemical reaction are represented in terms of swarms of independent trajectories that classically evolve on the support of the corresponding electronic potential energy surfaces. Such adiabatic, or Born-Oppenheimer, surfaces are calculated as the eigenvalues of the electronic time-independent Schr\"odinger equation at given nuclear configurations. The key nonadiabatic effects, beyond the Born-Oppenheimer approximation, i.e., transitions between electronic states, are mimicked by stochastic hops of the trajectories from one potential energy surface to another, usually occurring in localized regions of configuration space. Each trajectory of the swarm is coupled to the quantum-mechanical evolution of the electronic time-dependent state -- even though not all surface hopping implementations require this.\cite{Slavicek_JCTC2020, Agostini_JCTC2021}

Surface hopping has weaknesses, which have been reported extensively in the literature: it suffers from overcoherence, due to the disconnect between the dynamics of the classical nuclei and of the quantum electrons\cite{Granucci_JCP2007, Yang_JCP2011, Shenvi_JCP2011, Zoccante_JCP2010, Subotnik_ARPC2016, Prezhdo_JCP2012, PERSICO2024273, Martens_2019, Truhlar_JCP2004, Truhlar_JCTC2023}; when hops occur, energy conservation for the hopping trajectory is enforced, which causes frustrated hops if a trajectory hopping to a state of higher potential energy does not have enough kinetic energy to compensate for the increase in potential\cite{gomez2024benchmarking}; the velocity rescaling method applied to enforce energy conservation after the hops can significantly affect the system's dynamics;\cite{plasser2019strong,vindel2021study,Curchod_PCCP2020,toldo2024recommendations,dupuy2024exact, Sangiogo-Gil2025} 
using a swarm of independent trajectories to mimic the evolution of a quantum wavepacket misses non-local quantum-mechanical effects\cite{Shenvi_JCP2011, martens2016surface, Agostini_JCTC2021, Agostini_JCTC2024,subotnik2016understanding}. These weaknesses are consequences of the fact that surface hopping was not deduced from first principles, even though derivations~\cite{amati2023detailed, martens2016surface, subotnik2013can} and qualitative justifications~\cite{Gross_PRL2013, Gross_JCP2015} were proposed later on. Despite these weaknesses, as said above, surface hopping is a very powerful tool to model time-resolved photochemistry, as a recent Prediction Challenge~\cite{cyclobutanone_challenge} demonstrated. Outside of its ``comfort zone'', however, surface hopping can perform poorly in its standard formulation, and care must be taken to assess possible extensions either of the theory or of the applications, for instance when dealing with external time-dependent fields,~\cite{bajo2014interplay, Curchod_JPCA2019} like laser pulses, and with the creation of electronic wavepackets.~\cite{Agostini_JCP2024}. In the literature, these situations are more and more often studied with nonadiabatic dynamics simulations as the fields of femtochemistry and attochemistry continue to evolve on the experimental side,~\cite{Fert2024, Bergmann2021} which motivates the interest in continuing exploring alternative approaches to simulate nonadiabatic molecular dynamics.

Curing or alleviating the weaknesses of surface hopping remains a challenging and interesting problem from the theoretical viewpoint, especially if grounded on a rigorous quantum-mechanical basis, which might open the possibility of achieving a robust algorithm for extensions and generalizations. It is with this perspective that surface-hopping-based techniques continue to appear in the literature, as for instance consensus surface hopping (CSH),~\cite{martens2016surface} surface hopping based on the exact factorization (SH-XF),~\cite{Min_JPCL2018} coupled-trajectory Tully surface hopping (CT-TSH),~\cite{Agostini_JCTC2021} the mapping approach to surface hopping (MASH),~\cite{Richardson_JCP2023} quantum-trajectory surface hopping based on the exact factorization (QTSH-XF),~\cite{Maitra_JPCL2024} quantum-trajectory surface hopping with simplified decay of mixing (QTSH-SDM).~\cite{Akimov_JCTC2025} Motivated by these ongoing efforts and the clear interest of the physical-chemistry community in testing and benchmarking~\cite{RoadMap2025}
nonadiabatic molecular dynamics methods, we propose here some ideas based on the exact factorization~\cite{Gross_PRL2010, Agostini_PCCP2024, Gross_PRL2015, Gross_JPCL2017} to tackle the weaknesses of surface hopping. Specifically, we aim to address the following question: What is missing in surface hopping that would cure all its weaknesses at once? We believe that the idea of \textsl{coupled trajectories} emanating from the exact factorization is capable of indicating the route to follow. Specifically, this route implies that the ``surface hopping'' scheme is only introduced as a tool to simplify conceptually and numerically the evolution equations derived from the exact factorization, while the fundamental issues responsible for decoherence -- and the resulting internal inconsistency -- are bypassed by the coupling among the trajectories.
Note that a coupled-trajectory-based surface hopping scheme has been developed previously~\cite{Agostini_JCTC2021, Agostini_JCTC2024}, but, as we will show below, it inherits some of the issues of the original surface hopping since the idea of coupled trajectories was not fully ``embraced'' by all aspects of the algorithm. Instead, our new proposition appears to be more robust and stable, and, thus, we believe that our strategy is an important step forward in the development of novel ideas for nonadiabatic molecular dynamics schemes. It is worth mentioning here that other methodologies exist that are based on coupled trajectories, as for instance full and ab initio multiple spawning (FMS and AIMS),~\cite{Curchod2018} ab initio multiple cloning (AIMC),~\cite{Makhov2014,Freixas2018,Song2021} direct-dynamics variational multi-configurational Gaussian (DD-vMCG) method.~\cite{Worth2020} In these cases, the coupling among the trajectories emerges from the underlying representation of the nuclear wavepackets associated to the electronic states, that are expressed as linear combinations of moving Gaussians, usually referred to as trajectory basis functions. 

To present and to test our new algorithm, the remainder of the paper is structured as follows. First, our ideas of surface-hopping-based algorithms with coupled trajectories derived from the exact factorization of the molecular wavefunction are presented in Section~\ref{sec: theory}. Then, Section~\ref{sec: models} assesses the performance of these algorithms on the full-dimensional linear vibronic coupling models for fulvene and for 4-(dimethyloamino)benzonitrile (DMABN),~\cite{gomez2024benchmarking} whose molecular counterparts, together with ethylene, have been recently proposed as ``molecular Tully models'' by Ibele and Curchod.~\cite{Curchod_PCCP2020} Our conclusions are stated in Section~\ref{sec: conclusions}.

\section{Coupled-trajectory methods combined with surface hopping}\label{sec: theory}
The exact factorization (EF)~\cite{Gross_PRL2010, Agostini_PCCP2024} proposes to factor the time-dependent molecular wavefunction $\Psi(\mathbf r,\mathbf R,t)$ as the product of a marginal nuclear amplitude $\chi(\mathbf R,t)$, or wavefunction, and a conditional electronic amplitude $\Phi(\mathbf r,t;\mathbf R)$, 
\begin{align}\label{eqn: EF}
\Psi(\mathbf r,\mathbf R,t) = \chi(\mathbf R,t)\Phi(\mathbf r,t;\mathbf R)
\end{align}
thus yielding the molecular time-dependent Schr\"odinger equation (TDSE) 
\begin{align}\label{eqn: TDSE}
i\hbar\frac{\partial}{\partial t}\Psi(\mathbf r,\mathbf R,t) = \left[\sum_{\nu=1}^{N_n} \frac{-\hbar^2\boldsymbol\nabla_\nu^2}{2M_\nu}+\hat H_{el}(\mathbf r,\mathbf R)\right]\Psi(\mathbf r,\mathbf R,t)
\end{align}
in terms of two coupled evolution equations: an effective nuclear TDSE and an electronic evolution equation, 
\begin{align}
i\hbar\frac{\partial}{\partial t}\chi(\mathbf R,t) &= \left[\sum_{\nu=1}^{N_n} \frac{[-i\hbar\boldsymbol\nabla_\nu+\mathbf A_\nu(\mathbf R,t)]^2}{2M_\nu}+\epsilon(\mathbf R,t)\right] \chi(\mathbf R,t) \label{eqn: n eqn}\\
i\hbar\frac{\partial}{\partial t}\Phi(\mathbf r,t;\mathbf R) &= \left[\hat H_{el}(\mathbf r,\mathbf R)+\hat U_{en}[\Phi,\chi](\mathbf R,t)-\epsilon(\mathbf R,t)\right]\Phi(\mathbf r,t;\mathbf R) \label{eqn: el eqn}
\end{align}
In the previous equations, we used the following notation: $\mathbf r,\mathbf R$ indicate the electronic and nuclear configuration space variables, respectively; the molecular Hamiltonian in Eq.~(\ref{eqn: TDSE}) is the sum of the nuclear kinetic energy and the electronic Hamiltonian $\hat H_{el}(\mathbf r,\mathbf R)$; the index $\nu$ labels the nuclei; $M_\nu$ is the mass of the nucleus $\nu$.

The nuclear dynamics expressed in Eq.~(\ref{eqn: n eqn}) is generated by a time-dependent vector potential $\mathbf A_\nu(\mathbf R,t) = \langle\Phi(t;\mathbf R)| -i\hbar\boldsymbol\nabla_\nu\Phi(t;\mathbf R) \rangle_{\mathbf r}$ and a time-dependent potential energy surface $\epsilon(\mathbf R,t) = \langle\Phi(t;\mathbf R)| \hat H_{el} + \hat U_{en} -i\hbar\partial_t |\Phi(t;\mathbf R) \rangle_{\mathbf r}$, reminiscent of classical electromagnetic potentials and representing the effect of the electrons. Formally, an elegant and rigorous procedure can be introduced to derived classical-like Hamilton's equations from this effective TDSE~(\ref{eqn: n eqn}) to generate nuclear trajectories under the effect of the force determined from $\mathbf A_\nu(\mathbf R,t)$ and $\epsilon(\mathbf R,t)$.\cite{Ciccotti_EPJB2018} In turn, the nuclei affect the electronic dynamics entering the evolution equation with their momentum/velocity field and with their spatial distribution encoded in the electron-nuclear coupling operator $\hat U_{en}= \sum_\nu [-i\boldsymbol\nabla_\nu-\mathbf A_\nu(\mathbf R,t)]^2/(2M_\nu)+(-i\hbar\boldsymbol\nabla_\nu\chi(\mathbf R,t)/\chi(\mathbf R,t)+\mathbf A_\nu(\mathbf R,t))\cdot(-i\hbar\boldsymbol\nabla_\nu -\mathbf A_\nu(\mathbf R,t))/M_\nu$. Specifically, using the polar representation of the nuclear wavefunction, i.e., expressing it in terms of its modulus $|\chi|$ and phase $S$, we can rewrite the leading~\cite{AgostiniEich_JCP2016} term\footnote{Leading order in terms of the electron-nuclear mass ratio.} of the electron-nuclear coupling operator as
\begin{align}
\frac{-i\hbar\boldsymbol\nabla_\nu\chi(\mathbf R,t)}{\chi(\mathbf R,t)}+\mathbf A_\nu(\mathbf R,t) &= \left[\boldsymbol\nabla_\nu S(\mathbf R,t)+\mathbf A_\nu(\mathbf R,t)\right] +i \frac{-\hbar\boldsymbol\nabla_\nu |\chi(\mathbf R,t)|^2}{2|\chi(\mathbf R,t)|^2} \label{eqn: qmom 1}\\
&=\mathbf P_\nu(\mathbf R,t) + i\mathbf Q_\nu(\mathbf R,t)\label{eqn: qmom 2}
\end{align}
The term in square brackets on the right-hand side of Eq.~(\ref{eqn: qmom 1}) defines the nuclear momentum field $\mathbf P_\nu(\mathbf R,t)$, including the effect of the time-dependent vector potential, while the quantum momentum $\mathbf Q_\nu(\mathbf R,t)$ appears as an imaginary correction to the real momentum field and depends on the spatial nuclear distribution (the second term on the right-hand side of Eq.~(\ref{eqn: qmom 1})). If the quantum momentum is neglected in Eq.~(\ref{eqn: qmom 2}) and in the expression of the electron-nuclear coupling operator, the classical limit of the EF equations leads to Ehrenfest-like equations,~\cite{Gross_JCP2014} and, specifically, the effect of the nuclei on the electronic evolution is simply encoded in the momentum. Equation~(\ref{eqn: el eqn}) would then yield a ``standard'' electronic dynamics as for instance in the original surface hopping~\cite{Tully_JCP1990} scheme. Instead, the presence of the quantum momentum enables to track the delocalization of the nuclear distribution in time and to transfer this information to the electronic dynamics, thus naturally providing a channel for decoherence effects, as discussed abundantly in the literature.~\cite{Agostini_EPJB2018, Gross_PRL2015, Gross_JPCL2017} Interestingly, the quantum momentum is reminiscent of a coherence-damping term that appears in the evolution equation for the phase-space-dependent elements of the density matrix in QTSH-SDM~\cite{Akimov_JCTC2025} where auxiliary Gaussians in phase space are introduced.

As simple as this, the coupled-trajectory idea emerges in EF: while the momentum field can be interpreted classically, as it is a local property that can be encoded in a trajectory, the very nature of the spatial nuclear distribution is non-local. Therefore, to determine the quantum momentum, the trajectories become coupled since only the whole ensemble yields the necessary non-local information. Extensive analyses of the time-dependent potentials of the EF have justified some additional -- necessary for computational efficiency -- approximations that ultimately led to the trajectory-based quantum-classical algorithm dubbed \textsl{coupled-trajectory mixed quantum-classical} (CT-MQC). These approximations have been extensively discussed in the literature~\cite{Gross_JCTC2016, Agostini_JCTC2020_1, Agostini_JCTC2021, Agostini_EPJB2021}. CT-MQC equations express the electronic evolution and the classical-like nuclear forces in the adiabatic basis, formed by the set of the eigenstates of the electronic Hamiltonian $\hat H_{el}$, and it requires to calculate at every time step of dynamics the $3 N_n-$dimensional nonadiabatic coupling vectors (NACVs) for each pair of electronic states considered in the simulation. In order to circumvent this costly operation, additional simplifications have been considered, namely (i) to use purely adiabatic forces to drive the nuclear dynamics, since the time-dependent scalar potential reproduces piecewise the adiabatic shapes~\cite{Gross_PRL2013}, and (ii) to employ the idea of hopping with associated velocity rescaling, to compensate for the effect of the time-dependent vector potential that couples to the nuclear kinetic energy in the nuclear TDSE~\cite{Gross_JCP2015, Agostini_JPCL2017}. Based on these observations, the CT-TSH algorithm was proposed~\cite{Agostini_JCTC2021, Agostini_JCTC2024}.

CT-TSH is capable to circumvent the issue of overcoherence via the presence of the quantum momentum in the electronic evolution, which encodes, as argued above, information about the nuclear spatial delocalization. Nonetheless, as we will show below and as previously observed~\cite{Agostini_MP2024}, CT-TSH suffers from frustrated hops since, apart from the evaluation of the quantum momentum, the trajectories behave as independent entities and their coupling is not fully exploited, in particular to enforce energy conservation after the hops. Instead, by simply understanding that the trajectories need to cooperate to yield a meaningful approximation of the nuclear quantum dynamics, the additional issues discussed above can be easily cured: if the trajectories share their kinetic energy when hops occur, so as to conserve the total energy of the swarm, frustrated hops are avoided (or strongly reduced to a statistically irrelevant amount) and different velocity rescaling schemes do not affect the outcome of the simulation. Note that the idea of energy conservation over the full ensemble of coupled trajectories has been previously suggested by Villaseco Arribas and Maitra~\cite{VM23, Agostini_JCTC2023} in the context of CT-MQC.

The CT-TSH algorithm~\cite{Agostini_JCTC2021, Agostini_JCTC2024} is thus summarized with the following equations
\begin{align}
\dot C_{I}^{\alpha}(t) =& \dot C_{I,\textrm{TSH}}^{\alpha}(t)+\dot C_{I,\textrm{CT}}^{\alpha}(t) \label{eqn: Cdot}\\
\mathbf F_\nu^\alpha(t)=& -\boldsymbol\nabla_\nu E_{\textrm{active}}^\alpha\label{eqn: adia force}
\end{align}
We used the symbol $C_{I}^{\alpha}(t)$ for the coefficient of the electronic state $I$ in the expansion of the electronic time-dependent conditional amplitude of Eq.~(\ref{eqn: EF}) in the adiabatic basis. The trajectories are labeled with the index $\alpha$, thus the dependence on $\mathbf R\rightarrow \mathbf R^\alpha(t)$ is simply indicated via the superscript $\alpha$. The electronic coefficients evolve according to a standard ``Tully surface hopping'' (TSH) term~\cite{Tully_JCP1990}
\begin{align}\label{eqn: Cdot TSH}
\dot C_{I,\textrm{TSH}}^{\alpha}(t)=-\frac{i}{\hbar}E_I^\alpha C_I^\alpha(t)-\sum_J\sum_{\nu=1}^{N_n}\boldsymbol v_\nu^\alpha(t) \cdot \mathbf d_{\nu,IJ}^\alpha C_J^\alpha(t)
\end{align}
and to a ``coupled-trajectory'' (CT) contribution that depends on the quantum momentum
\begin{align}\label{eqn: Cdot qmom}
\dot C_{I,\textrm{CT}}^{\alpha}(t)=\sum_{\nu=1}^{N_n}\frac{\mathbf Q_\nu^\alpha(t)}{\hbar M_\nu}\cdot\left(\mathbf f_{\nu,I}^\alpha-\sum_J \left|C_J^{\alpha}(t)\right|^2 \mathbf f_{\nu,J}^\alpha\right)C_I^\alpha(t)
\end{align}
In the above equations, we introduced the following symbols: the adiabatic energy of state $I$ at the position $\mathbf R^\alpha(t)$ as $E_I^\alpha$, the NACV between states $I$ and $J$ at the position $\mathbf  R^\alpha(t)$ as $\mathbf d_{\nu,IJ}^\alpha$, the adiabatic force of state $J$ accumulated over time along the trajectory $\mathbf R^\alpha(t)$ as $\mathbf f_{\nu,J}^\alpha = \int^t (-\boldsymbol\nabla_\nu E_J^\alpha)d\tau$, the velocity of the nucleus $\nu$ along the trajectory  $\mathbf R^\alpha(t)$ as $\boldsymbol v_\nu^\alpha(t)$.
In addition, the classical force in Eq.~(\ref{eqn: adia force}) driving the evolution of the nuclei is purely adiabatic and is given by the gradient of the active state. In CT-TSH, the active state is identified at each time step according to the fewest-switches procedure~\cite{Tully_JCP1990}.

In CT-TSH, the coupling among the trajectories is only employed to determine the quantum momentum in $\dot C_{I,\textrm{CT}}^{\alpha}(t)$. This is done, as in standard CT-MQC, by reconstructing the nuclear density at each time step as a sum of frozen Gaussians centered at the positions of the trajectories, as abundantly documented in Refs.~[\!\!\citenum{Maitra_JCTC2018, Maitra_M2022, Agostini_JCTC2021, Agostini_JCTC2020_1}]. The use of frozen Gaussians to reconstruct the nuclear density yields an analytical expression for the quantum momentum. However, this form violates the physical conditions that no population transfer between two electronic states should occur when averaged over the ensemble of trajectories if the NACV between them is zero. This condition needs, thus, to be imposed a posteriori, by introducing a modified expression of the quantum momentum at each time step.~\cite{Gossel_JCP2019,Maitra_M2022} Understanding the importance of fully exploiting the coupling among the trajectories in surface hopping leads to the derivation of the combined coupled-trajectory Tully surface hopping (CCT-TSH), that we illustrate below and that we test in Section~\ref{sec: models}.

\begin{figure}
    \centering
    \includegraphics[width=.9\linewidth]{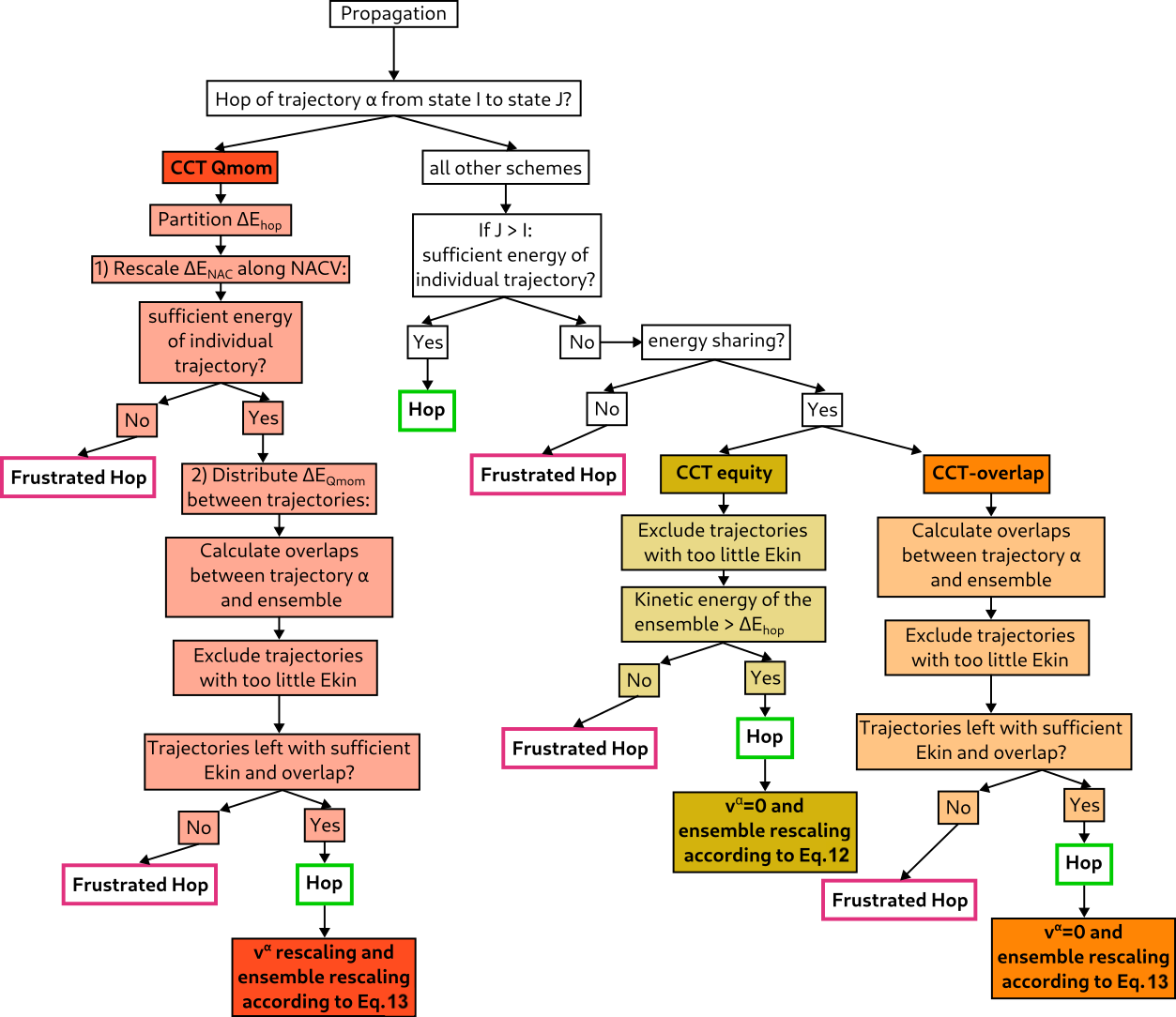}
    \caption{Schematic overview of the CCT-TSH algorithm.}
    \label{fig:scheme}
\end{figure}

CCT-TSH is based on the idea of ``energy sharing'' to conserve the energy of the swarm of trajectories and to alleviate the frustrated hops. CCT-TSH is based on Eqs.~(\ref{eqn: Cdot}) and~(\ref{eqn: adia force}), and we introduce an energy-sharing procedure only at the moment of the hops. We propose \textit{three} schemes for the energy sharing, which are schematically summarized in Fig.~\ref{fig:scheme}: (i) the equity-based scheme, (ii) the overlap-based scheme, and (ii) the quantum-momentum-based scheme. We stress here that, while the idea of energy sharing appears well-founded from a quantum-mechanical viewpoint, its practical implementation in a trajectory-based scheme is quite arbitrary. Therefore, the schemes proposed here are initial attempts to introduce this idea in CCT-TSH and we expect them to evolve based on future additional tests and benchmarks.

Before detailing the energy-sharing schemes, let us introduce the following quantities. The trajectories are labeled with Greek indices, and the hopping trajectory is the trajectory $\alpha$, with kinetic energy indicated with the symbol $E_{\text{kin}}^\alpha$. The variation in potential energy for a hop from state $I$ towards a state $J$ of higher energy is $\Delta E^\alpha_{\text{hop}}<0$. The Cartesian components of the vectors are indicated with Latin indices, thus for instance the symbol $v_i^{\beta}(t)$ stands for the component $i$ of the velocity of the trajectory $\beta$ at time $t$. For scheme (ii), we define the overlap between two trajectories, i.e., $S_{\alpha\beta}$, as the overlap between two Gaussians centered at the positions of the trajectories $\alpha$ and $\beta$. Note that the algorithm does not depend on the value of the width of the Gaussians, as long as they are the same for all trajectories and are large enough to guarantee accounting for all potentially contributing trajectories, since $S_{\alpha\beta}$ is only used to quantify the proximity of the trajectories. In addition, since the trajectories are used to reconstruct the quantum nuclear density in configuration space, we chose to consider their overlap in configuration space and not in phase space (indeed, alternative choices might be equally or even better performing). For scheme (iii), we identify two channels for energy sharing, one driven by the nonadiabatic couplings between states $I$ and $J$, $\boldsymbol d_{\nu,IJ}^\alpha$, and one driven by the quantum momentum, $\mathbf Q_\nu^{\alpha}$, both evaluated at the position of the hopping trajectory.
Therefore, we use the scalar quantities $\boldsymbol v^\alpha \cdot \boldsymbol d_{IJ}^\alpha= \sum_\nu \boldsymbol v_\nu^\alpha  \cdot  \boldsymbol d_{\nu,IJ}^\alpha $ and $\boldsymbol q^{\alpha} \cdot \boldsymbol f_{IJ}^\alpha = \hbar^{-1}\sum_\nu(\mathbf Q_\nu^{\alpha}/M_\nu) \cdot (\mathbf f_{\nu,I}^\alpha-  \mathbf f_{\nu,J}^\alpha)) \textrm{Re}[\bar C_J^\alpha(t) C_I^\alpha(t)]$, to decompose the energy variation to be compensated for the hop to occur.
As described above, these scalar quantities appear in the electronic evolution equation of CT-TSH and CCT-TSH, with $\boldsymbol f_{\nu,IJ}^\alpha =  \hbar^{-1}\textrm{Re}[\bar C_J^\alpha(t) C_I^\alpha(t)](\mathbf f_{\nu,I}^\alpha -\mathbf f_{\nu,J}^\alpha)$ the coefficient-weighted, time-integrated difference of the adiabatic forces for states $I$ and $J$ and with $\boldsymbol q^\alpha_\nu=\boldsymbol Q_\nu^{\alpha}/M_\nu$ a ``pseudo-velocity''. Note that if Eq.~(\ref{eqn: Cdot qmom}) is rewritten as
\begin{align}\label{eqn: Cdot qmom rewritten}
\dot C_{I,\textrm{CT}}^{\alpha}(t)=\sum_J\left[\sum_{\nu=1}^{N_n}\frac{\mathbf Q_\nu^\alpha(t)}{\hbar M_\nu}\cdot\left(\mathbf f_{\nu,I}^\alpha-  \mathbf f_{\nu,J}^\alpha\right)\bar C_J^\alpha(t)C_I^\alpha(t)\right]C_J^\alpha(t)
\end{align}
this expression of $\dot C_{I,\textrm{CT}}^{\alpha}(t)$ is formally similar to the second term on the right-hand side of $\dot C_{I,\textrm{TSH}}^{\alpha}(t)$ in Eq.~(\ref{eqn: Cdot TSH}) and justifies the use of the (real part of the) term in square bracket to evaluate the energy sharing.

Using the definitions given above, the energy-sharing schemes are as follows:
\begin{enumerate}[(i)]
\item \textbf{Equity-based.} If the trajectory $\alpha$ has insufficient kinetic energy to compensate a hop from state $I$ to a higher-lying state $J$, the difference $\Delta E_{\text{hop}}^{\text{CCT}} =\Delta E^\alpha_{\text{hop}} - E_{\text{kin}}^\alpha$ is distributed among all remaining trajectories of the swarm depending on their kinetic energy, but we defined a kinetic energy threshold below which a trajectory is not considered for the energy sharing. If the sum of the kinetic energies of the contributing trajectories is sufficient for the hop, their velocities $v_i^{\beta}$ are rescaled as 
\begin{equation}
{v_i^{\beta}}^\prime=v_i^\beta  \sqrt{ 1+ \frac{\Delta E_{\text{hop}}^{\text{CCT}}}{\sum_\gamma E_{\text{kin}}^\gamma}}
\end{equation}
where ${v_i^{\beta}}^\prime$ is the velocity after rescaling. Weighting the contribution of the trajectories $\beta$ according to their kinetic energy ensures a minimal impact on the trajectory itself. Since the trajectory $\alpha$ uses its entire kinetic energy for the hop, its new velocity after the hop is $\boldsymbol{v}^\alpha=0$.
\item \textbf{Overlap-based.} If the trajectory $\alpha$ has insufficient kinetic energy to compensate a hop from state $I$ to a higher-lying state $J$, the contribution of the trajectories $\beta$ is determined based on their spatial proximity to the trajectory $\alpha$. The velocity rescaling is done as
\begin{equation}
    {v_i^{\beta}}^\prime=v_i^\beta  \sqrt{1 + \frac{\Delta E_{\text{hop}}^{\text{CCT}} S_{\beta\alpha}}{E_{\text{kin}}^\beta\sum_\gamma S_{\gamma\alpha} }}
\end{equation}
where $\sum_\gamma S_{\gamma\alpha}$ is only summed over trajectories whose kinetic energy is large enough for the rescaling. The trajectory $\alpha$ uses its entire kinetic energy for the hop, and its new velocity after the hop is $\boldsymbol{v}^\alpha=0$.
\item \textbf{Quantum-momentum-based.} The amount of energy to be rescaled $\Delta E^{\alpha}_{\text{hop}}$ is divided in two contributions: One contribution, $\Delta E^{\alpha}_{\text{NAC}}$, that will be compensated by the kinetic energy of the hopping trajectory $\alpha$ and that will be rescaled along the NACV, and a second contribution, $\Delta E^{\alpha}_{\text{Qmom}}$, that will be compensated by the kinetic energy of the whole swarm of trajectories based on (ii). The two contributions are calculated according to
\begin{align}\label{eqn: qmom ES nac}
\Delta E^{\alpha}_{\text{NAC}} = \Delta E^{\alpha}_{\text{hop}}\frac{| \boldsymbol v^\alpha \cdot \boldsymbol  d_{IJ}^\alpha|}{| \boldsymbol v^\alpha \cdot \boldsymbol  d_{IJ}^\alpha| +| \boldsymbol q^{\alpha} \cdot \boldsymbol f_{IJ}^\alpha|}
\end{align}
and
\begin{align}\label{eqn: qmom ES qmom}
\Delta E^{\alpha}_{\text{Qmom}} = \Delta E^{\alpha}_{\text{hop}}\frac{|\boldsymbol q^{\alpha} \cdot \boldsymbol f_{IJ}^\alpha|}{| \boldsymbol v^\alpha \cdot \boldsymbol  d_{IJ}^\alpha| +| \boldsymbol q^{\alpha} \cdot \boldsymbol f_{IJ}^\alpha|}
\end{align}
such that $\Delta E^{\alpha}_{\text{hop}} = \Delta E^{\alpha}_{\text{NAC}} + \Delta E^{\alpha}_{\text{Qmom}}$.
\end{enumerate}
In general, in CCT-TSH, we assume that the kinetic energy of the hopping trajectory $\alpha$ is fully used first, before requesting the cooperation of the surrounding trajectories via the energy sharing. Therefore, the velocity rescaling options used in the equity-based and in the overlap-based sharing schemes are either isotropic, meaning that all components of the velocity vector are equally rescaled, or hybrid, meaning that the component of the velocity along the NACV is rescaled if possible, and otherwise it is isotropic. Nonetheless, we will show that the velocity rescaling procedure does not impact the results. As indicated above, in all energy-sharing options, the contributing trajectories are selected based on their kinetic energy, which needs to be above a certain threshold. This option is considered mainly to avoid that such contributing trajectories suffer later on in the dynamics from the problem of frustrated hops.

It is worth discussing at this point the decoherence problem of surface hopping~\cite{Subotnik_ARPC2016, PERSICO2024273} and its relation to the internal consistency. Internal consistency refers to the agreement between the two ways of estimating the population of the electronic states in surface hopping: the fraction of trajectories associated to, or evolving in, an electronic state (F); the trajectory-averaged population propagated along each trajectory (P), since usually a quantum-dynamical evolution equation for the electronic state is integrated along each trajectory. However, in general and unless decoherence corrections are used, it cannot be ensured that the two estimates agree with each other.
The original hopping algorithm~\cite{tully1971trajectory} ensures this internal consistency by using an average hopping probability over the swarm of trajectories for which the two estimates for the populations are the same. However, as soon as an independent trajectory formalism is invoked and the hopping probability is calculated at the level of a single trajectory, internal consistency is no longer guaranteed. The coupled-trajectory schemes that we propose here are naturally designed to reintroduce the idea of an average hopping probability to restore internal consistency. However, within CT-TSH this might introduce unphysical artifacts. Such an average hopping algorithm will assign the same hopping probability to all trajectories regardless of their position. While this might ensure a correct distribution of the wavepacket/swarm with respect to the electronic states, it neglects the locality of the nonadiabatic couplings and the spatial distribution of the wavepacket, causing trajectories to undergo hops even in regions of negligible nonadiabatic coupling. This limitation has previously been overcome through the CSH algorithm,~\cite{martens2016surface} where an ensemble of stochastic trajectories is used to solve the semiclassical Liouville equation.~\cite{Martens1997,Donoso1998,Donoso2000,Donoso2000-2,Donoso2002,Kapral1999,Hanna2005,Ando2002,Ando2003,Nielsen2000} Note that in CSH, the velocity rescaling procedure is circumvented, since the total energy is naturally conserved over the entire ensemble of trajectories thanks to the underlying structure based on the Liouville equation. Therefore, while CCT-TSH might seem similar to CSH, it is fundamentally a different approach. 

For CCT-TSH, we propose an alternative approach for the hopping algorithm, namely to forgo the stochastic scheme and to introduce a deterministic hopping, as also done in MASH.~\cite{Richardson_JCP2023} Since the electronic evolution is derived from the EF equation for the conditional amplitude~(\ref{eqn: el eqn}), we drive the trajectory hops based on the values of the electronic populations predicted by this equation. In a way, the nuclear trajectories in CCT-TSH can be regarded merely as a support on which the time-dependent electronic state and the quantum momentum are evaluated. Therefore, it is crucial that the trajectories evolve following the electronic populations. For this reason, in CCT-TSH, the trajectories evolve in the electronic state with the largest population. Nonetheless, we will assess the performance of this ``deterministic'' version of CCT-TSH by comparing it to its fewest-switches version, where the active state is, instead, chosen as in CT-TSH.

\section{Fulvene and DMABN dynamics with coupled trajectories}\label{sec: models}
The CCT-TSH algorithms illustrated above are tested on the 30-dimensional linear vibronic coupling (LVC) model of fulvene and on the 57-dimensional LVC model of DMABN.\cite{gomez2024benchmarking} 

Upon photoexcitation to the first excited state S$_1$, fulvene decays through two conical intersections, the first one is a sloped conical intersection seam and is associated to a stretching of the ethenylic bond and the second one is a peaked conical intersection and is associated to the twist of the methylene group~\cite{Curchod_PCCP2020, mendive2010controlling}. This LVC model of fulvene reproduces only the sloped conical intersection, but not the peaked intersection. However, the latter does not appear to be important in the simulated short-time dynamics as argued previously~\cite{Alfalah2009,gomez2024benchmarking}. Furthermore, we are interested in correctly capturing the challenging dynamics at the slope conical intersection, where the wavepacket/swarm is partially reflected and thus crosses multiple times the nonadiabatic coupling region.  Also, note that the molecular system of fulvene has been shown to be highly sensitive to the parameters of the surface hopping algorithm, in terms of procedure to treat frustrated hops, velocity rescaling and time step.\cite{Curchod_PCCP2020,toldo2024recommendations,vindel2021study,Sangiogo-Gil2025} Therefore, it is a suitable test system to verify the robustness of the proposed CCT-TSH schemes. Concerning DMABN, the electronic structure and photodynamics of DMABN have been previously studied quite extensively in Refs.~[\!\!\citenum{Curchod2017, Du2015, Parusel2002, Rappoport2004, Curchod_PCCP2020, Worth_PCCP2023, Kelly_JPCL2024}]. For the LVC model proposed in Ref~.[\!\!\citenum{Worth_PCCP2023}], it was found that DMABN encounters an ``immediate'' conical intersection between states S$_1$ and S$_2$ after photoexcitation to S$_2$, which is accessible directly by the photoexcited wavepacket located in the Franck-Condon region. This leads to a fast population transfer from state S$_2$ to S$_1$. Ab initio studies have shown that DMABN has an additional conical intersection between S$_2$ and S$_1$ that is associated with a rotation of the dimethylamino group~\cite{Curchod2017}, but the LVC model does not account for this conical intersection.

\subsection{Computational details}
All algorithms employed for the simulations performed here are implemented in G-CTMQC~\cite{GCTMQC}, which is available open-source in GitLab. 

All simulations have been initialized with a swarm of 500 trajectories with positions and momenta Wigner sampled from the vibrational ground state of the electronic LVC ground state potential centered at the origin (in both position and momentum). The trajectories were propagated with a time step of 0.1~a.t.u. for 4200~a.t.u (about 100~fs of dynamics are simulated in this way). Such a small integration time step is necessary because the multiple time step procedure is currently being tested in G-CTMQC and it is, thus, not used in the calculations performed here. Some simulations have also been run by Wigner-sampling only the positions while initializing all the trajectories with the same zero momentum to assess the robustness of the algorithms with different sets of initial conditions.\footnote{Note that we expect the final results to be affected upon changing the initial conditions, since the solution of the ordinary differential equations representing the algorithms clearly depends on the initial conditions. However, our tests aim to assess the robustness of the energy-sharing procedures within different ways of initializing the dynamics.}

Numerical tests are performed using the original surface hopping algorithm (TSH) and the modified version that includes decoherence using the energy-based approach (TSH-ED) proposed by Granucci and Persico~\cite{Persico_JCP2007}. TSH and TSH-ED are based on Eqs.~(\ref{eqn: adia force}) and~(\ref{eqn: Cdot TSH}) combined with the fewest-switches procedure to select randomly the active state. In TSH-ED, the standard value of $0.1$~Ha is used for the energy parameter in the exponential decay applied to the electronic coefficients that are not associated to the active state. 

G-CTMQC allows to perform surface-hopping-based calculations using different schemes:
\begin{itemize}
\item for the treatment of frustrated hops (w/o inversion of the nuclear velocity for the trajectory experiencing a frustrated hop),
\item for the rescaling of the velocity after a hop (isotropical, along the NACVs, or a mixture of the two to avoid frustrated hops),
\item for the hopping procedure (either using fewest switches or by forcing the hop towards a state whose population reaches a certain threshold, which is chosen to be 0.5 in our calculations),
\item for the energy sharing (equity-based, overlap-based, quantum-momentum-based).
\end{itemize}
We provide as Supporting Information a script to help generate the input files to run all the calculations reported below, thus, all our data are easily reproducible. 

G-CTQMC is interfaced with QuantumModelLib~\cite{QML}, a library of analytical potentials that performs the diagonalization of the electronic LVC Hamiltonian
\begin{align}\label{eqn:el eqn for LVC}
\hat H_{el}(\mathbf q) = \hat{H}^{\mathrm{LVC}}(\mathbf q) -\left(\sum_{n=1}^{N_{\mathrm{modes}}}-\frac{\omega_n}{2}\frac{\partial^2}{\partial q_n^2}\right)\hat{I}_{N_{\mathrm{st}}\times N_{\mathrm{st}}},
\end{align}
to provide analytically adiabatic energies and gradients as well as the NACVs. Here, the second term on the right-hand side is the nuclear kinetic energy operator and the first term is the full
LVC Hamiltonian~\cite{Zobel2021, Zobel2023, Polonius2024, Domcke2012, Kppel2004, Ehrmaier2017,Green2022} 
\begin{align}
\hat{H}^{\mathrm{LVC}}(\mathbf q)=\hat H^{(0)}(\mathbf q)
+\hat{W}^{0}(\mathbf q)+\hat{W}^{1}(\mathbf q),
\end{align}
that has an analytical form, which (i) approximates harmonically the coupled (diabatic) potential energy surfaces in the vicinity of the Franck-Condon region (ii) including linear terms and (iii) coupling linearly the potentials associated to different electronic states. The LVC Hamiltonian is constructed in the diabatic representation, thus the electronic Hamiltonian is not diagonal in this basis, and the nuclear modes used are the normal modes $\mathbf q$ identified from the harmonic analysis of the ground-state potential energy surface in the Franck-Condon region. The zero-th order Hamiltonian is diagonal in the space of the electronic states
\begin{align}
\hat H^{(0)}(\mathbf q)=\left(\sum_{n=1}^{N_{modes}} -\frac{\omega_n}{2}\frac{\partial^2}{\partial q_n^2}+\frac{\omega_n}{2}q_n^2\right)\hat{I}_{N_{\mathrm{st}}\times N_{\mathrm{st}}},
\end{align}
with mass-frequency weighted normal coordinates $\mathbf q=\lbrace q_n\rbrace_{n=1,\ldots,N_{modes}}$ and harmonic frequencies $\omega_n$. For $N_{\mathrm{st}}$ electronic states, $\hat{W}^{0}(\mathbf q)$ and $\hat{W}^{1}(\mathbf q)$ are $N_{\mathrm{st}}\times N_{\mathrm{st}}$ matrices with elements
\begin{align}
{W}^{0}_{lm}=E_{l}\delta_{lm}
\end{align}
and
\begin{align}
{W}^{1}_{lm}(\mathbf q)=\begin{cases} \sum_n k^{(l)}_n q_n &l =m \\ \sum_n \lambda^{(l,m)}_n q_n  &l\neq m\end{cases}
\end{align}
The parameters $E_{l}$, $k^{(l)}_n$, $\lambda^{(l,m)}_n$, and $\omega_n$ used in our simulations are reported in Ref.~[\!\!\citenum{Worth_PCCP2023}]. 

\subsection{Numerical results}
\begin{figure}[htb!]
\includegraphics[width=.8\linewidth]{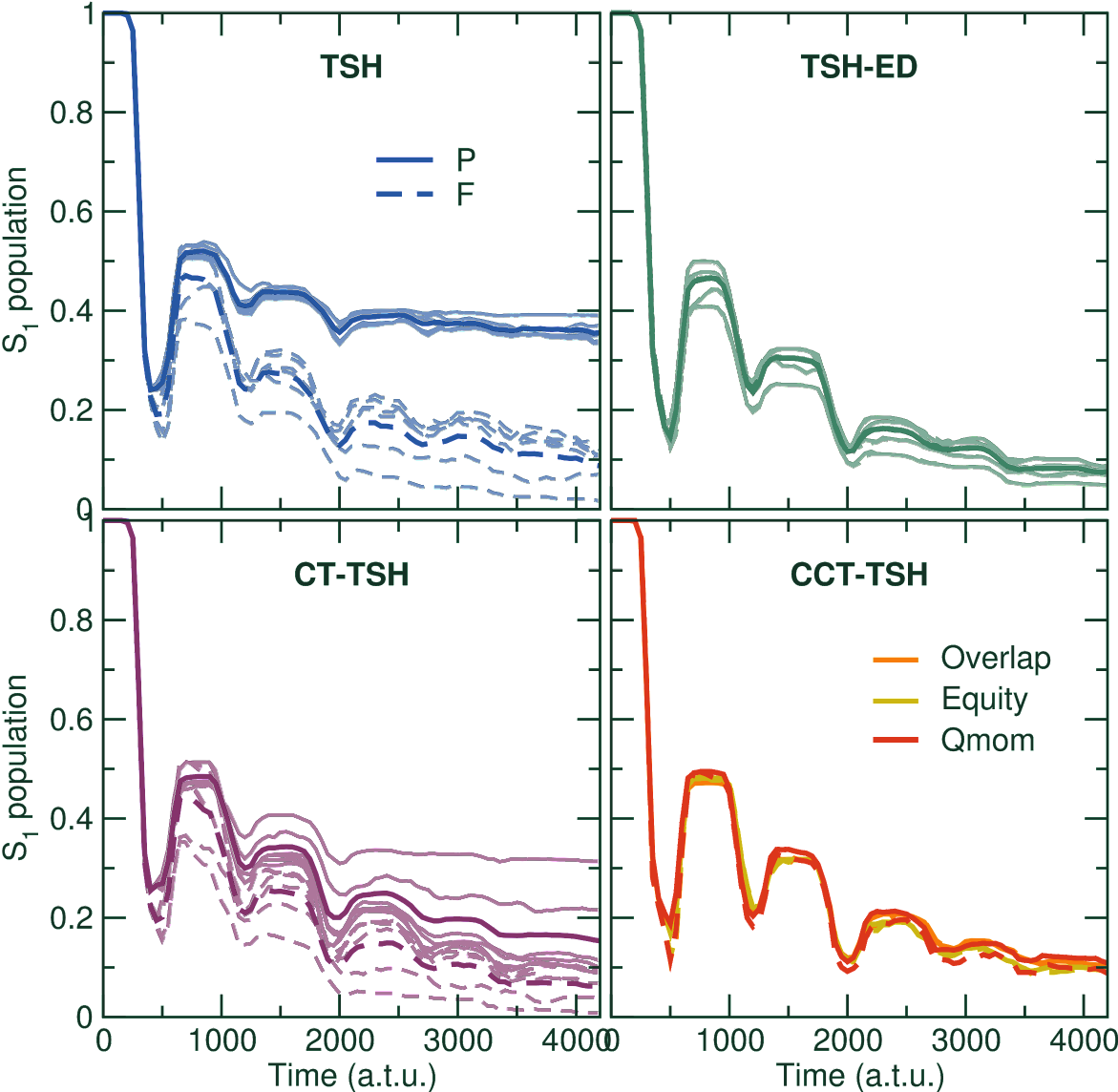}
\caption{Adiabatic population decays of S$_1$ in fulvene predicted by the different surface hopping schemes discussed in the main text. The lighter lines in each panel show the dynamics with different velocity rescaling procedures and treatments of frustrated hops, while the bold darker lines show the average of the individual runs. Solid lines show the average electronic populations (P) and dashed lines the fractions of trajectories (F).}
\label{fig:fulvene_pop}
\end{figure}

In Fig.~\ref{fig:fulvene_pop}, we show the adiabatic population curves of the fulvene LVC model upon excitation to the S$_1$ state. For the standard surface hopping approaches, namely for TSH, TSH-ED, CT-TSH, six different dynamics are averaged, using three different schemes to rescale the velocity after a hop, with each option used in combination with velocity inversion or not in the case of frustrated hops. The three schemes for the velocity rescaling are (1) isotropic, (2) along the NACVs, or (3) mixed, meaning that if (2) is not possible then (1). In a standard application of any surface hopping scheme to a molecular problem, each of these approaches would per se be regarded as a justified choice, which is why we chose to use the repetitions with different schemes to give us some information on the uncertainty of the results. In the case of CCT-TSH, in Fig.~\ref{fig:fulvene_pop}, we compare the results of the three energy-sharing schemes, i.e., equity-based, overlap-based, quantum-momentum-based, employed with the option of rescaling the velocity along the NACV and inverting the velocity in the case of frustrated hops. Note that using other velocity rescaling options does not change dramatically the results. In addition, we use a deterministic selection of the active state, depending on the state with largest populations.

Note that in Fig.~\ref{fig:fulvene_pop} we aim to point out the quite large distributions of values of the S$_1$ population, especially in TSH and CT-TSH, rather than the actual calculated values, which is why we only distinguish clearly the continuous lines -- electronic populations (P) determined from the electronic evolution equation --  and the dashed lines -- fractions of trajectories (F) evolving in S$_1$. The thick continuous and dashed lines, green for TSH, blue for TSH-ED and purple for CT-TSH, are the averages of the corresponding populations, with the spreading of the light-colored lines providing an idea of the associated uncertainty.

It can be clearly seen that for both TSH and CT-TSH internal consistency is not ensured, and in both cases, the electronic populations (P) decay slower than the fractions of trajectories (F). TSH-ED restores, as expected, almost perfect internal consistency. The population dynamics are very similarly predicted by all three methods, with a sharp decrease of S$_1$ population occurring after 250 a.t.u., but since the wavepacket/swarm of trajectories is partially reflected, we observe stepwise decays to around only 10\% of S$_1$ population at the end of the simulated dynamics. 
When looking at CCT-TSH, it can be observed that all three proposed schemes restore closely internal consistency and the population decays are almost identical with the three energy sharing schemes, which points to an encouraging robustness of the algorithm. However, the quantum-momentum-based sharing scheme yields a large number of frustrated hops (see Table~\ref{tab:frust_hops}). This result is hardly surprising for two reasons: the approximate nature of the quantum momentum, discussed above; the quantum-momentum-based energy sharing acts at every surface hop and not only when a hop would be frustrated on the single trajectory level. Note, in addition, that while the quantum-momentum-based energy sharing might seem the most physically appropriate, the choice to partition the energy according to Eqs.~(\ref{eqn: qmom ES nac}) and~(\ref{eqn: qmom ES qmom}) can be considered quite arbitrary and certainly different partitioning schemes are possible. For instance, in a case where the potential energy surfaces associated to two or more (coupled) electronic states remain parallel to each other (see Refs.~[\!\!\citenum{Liang2024,Suchan2025}] for discussions on the importance of these situations), Eq.~(\ref{eqn: qmom ES qmom}) would yield always zero, since the difference in gradients accumulated over time is zero as well. Similarly, the equity-based scheme might become ``unphysical'' if two or more branches of the nuclear density/trajectories travel far away from each other, thus, one would not expect plausible that all branches contribute to hops occurring locally in a region of a single branch.

The number of frustrated hops during the dynamics varies strongly between the different methods (see Table~\ref{tab:frust_hops}). As expected, when we allow only the component of the velocity along the NACVs to be rescaled, the number of frustrated hops is quite large in all methods. Applying a decoherence correction as in TSH-ED, not only improves the internal consistency but also strongly decreases the number of frustrated hops. If the full velocity vector can be rescaled isotropically, there are no (or a negligible number of) frustrated hops during the dynamics. 
The number of frustrated hops during the CT-TSH dynamics is comparable to TSH, but surprisingly much larger if only the component of the velocity along the NACVs can be rescaled. 
This issue is completely alleviated, albeit by construction, when using the equity-based and overlap-based CCT-TSH schemes. By contrast, the energy sharing scheme based on the quantum momentum shows a large number of frustrated hops. This can be explained by the modified hopping scheme where at each time step the energy sharing is attempted.  

\begin{table}[h]
    \centering
    \begin{tabular}{c|cccccc}
    \hline\hline
     & \multicolumn{6}{c}{\textbf{Number of frustrated hops}} \\
    \hline
        & \multicolumn{3}{c}{reflect frust hops} & \multicolumn{3}{c}{no reflection} \\
        &   full $v$ & mix & NACV & full $v$ & mix & NACV \\ 
        \hline
       TSH  & 81 & 89 & 490 & 92 & 103 & 591\\
       TSH-ED  & 0 & 1 & 147 & 0 & 1 & 130 \\
       CT-TSH & 57 & 119 & 1121 & 73 & 101 & 2049 \\
            & \multicolumn{2}{c}{Equity} & \multicolumn{2}{c}{Overlap} & \multicolumn{2}{c}{Qmom} \\
       CCT-TSH &\multicolumn{2}{c}{0} & \multicolumn{2}{c}{0} & \multicolumn{2}{c}{120}\\
       \hline\hline
    \end{tabular}
    \caption{Number of frustrated hops in the different surface hopping dynamics of fulvene.}
    \label{tab:frust_hops}
\end{table}

\begin{figure}[h]
\includegraphics[width=\linewidth]{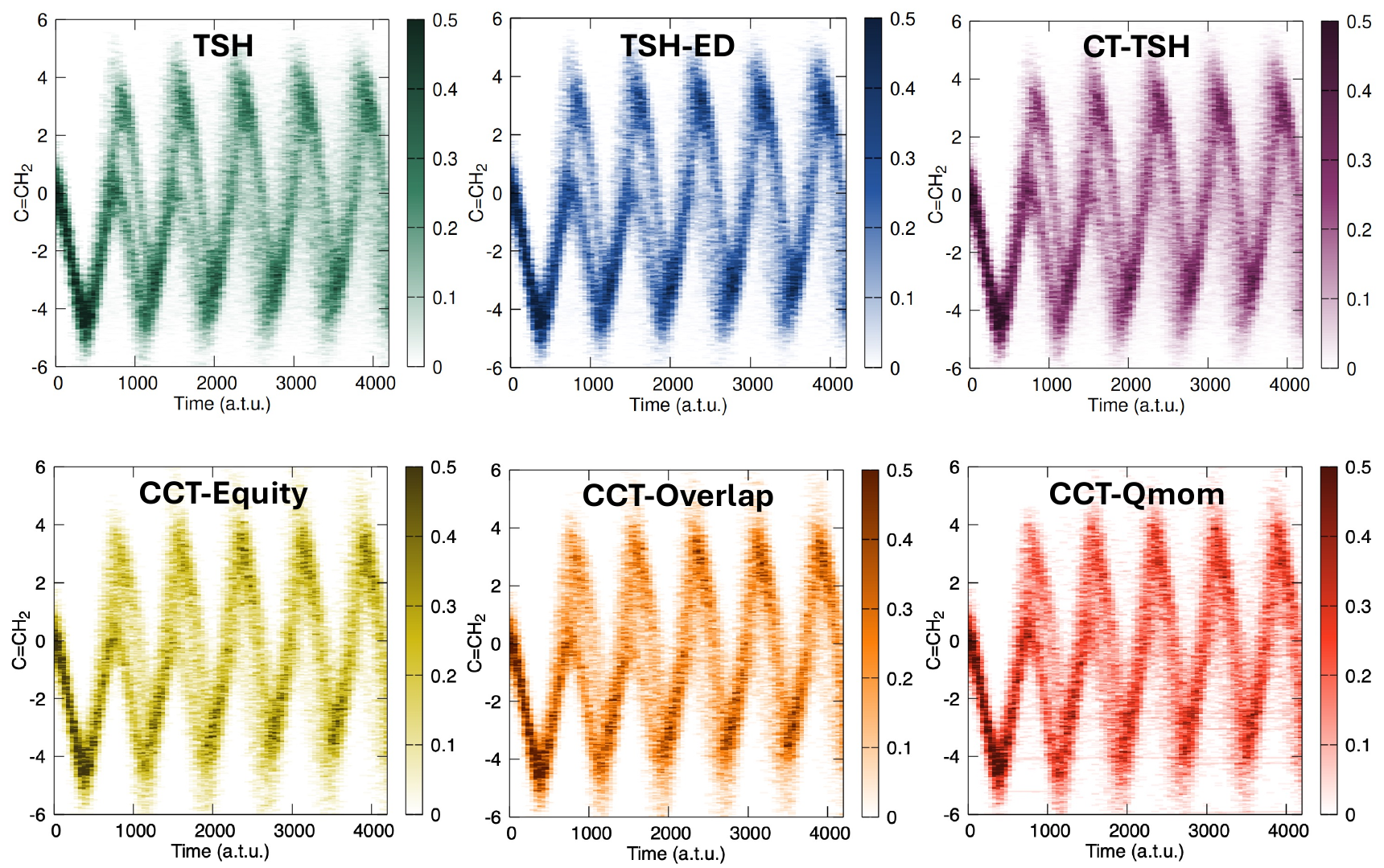}
\caption{Spatial evolution of the swarms of trajectories along the C=CH$_2$ stretching mode in fulvene. The $y$-axis is given in mass-weighted normal mode coordinates, negative values correspond to a stretching of the C=CH$_2$ bond, positive values to a contraction. TSH, TSH-ED and CT-TSH are averaged over the six different dynamics.}
\label{fig:v24}
\end{figure}

In order to ascertain that the energy sharing scheme does not lead to artifacts in the spatial distribution of the wavepacket, we report the distribution of the trajectories along the normal mode corresponding to the C=CH$_2$ stretching. This mode was chosen based on previous works where it has been observed that the sloped conical intersection between S$_1$ and S$_0$ is accessed via a stretching of the C=CH$_2$ bond.~\cite{mendive2010controlling,Worth_PCCP2023} Each trajectory was convoluted with a Gaussian of width 0.01 along this mode. We note that in Fig.~\ref{fig:v24} the $y$-axis is given in mass-weighted normal mode coordinates. Negative values correspond to an elongation of the bond, while positive values correspond to a contraction. 
The dynamics of TSH, TSH-ED and CT-TSH were averaged over all the six different runs.
Over time, the swarm of trajectories shows harmonic oscillations along this bond, initially elongating it, which drives the population decay. After the first elongation, around 500 to 1000 a.t.u., a splitting of the swarm of trajectories can be observed. One part contracts the bond further whereas another part contracts only until the equilibrium position before elongating again. This splitting can be attributed to the electronic population, where the part that decays to the ground state harmonically oscillates between $-4$ and $4$ and the part that gets reflected on the excited state oscillates only in the elongation between $0$ and $-4$. In all methods, TSH, TSH-ED and CT-TSH, these two branches are clearly distinguishable between 500 and 2000 a.t.u.
A similar behavior is exhibited by the CCT-TSH trajectories, and as with the population traces, their behavior seems to be in excellent agreement with TSH-ED results.

\begin{figure}[h]
\includegraphics[width=.6\linewidth]{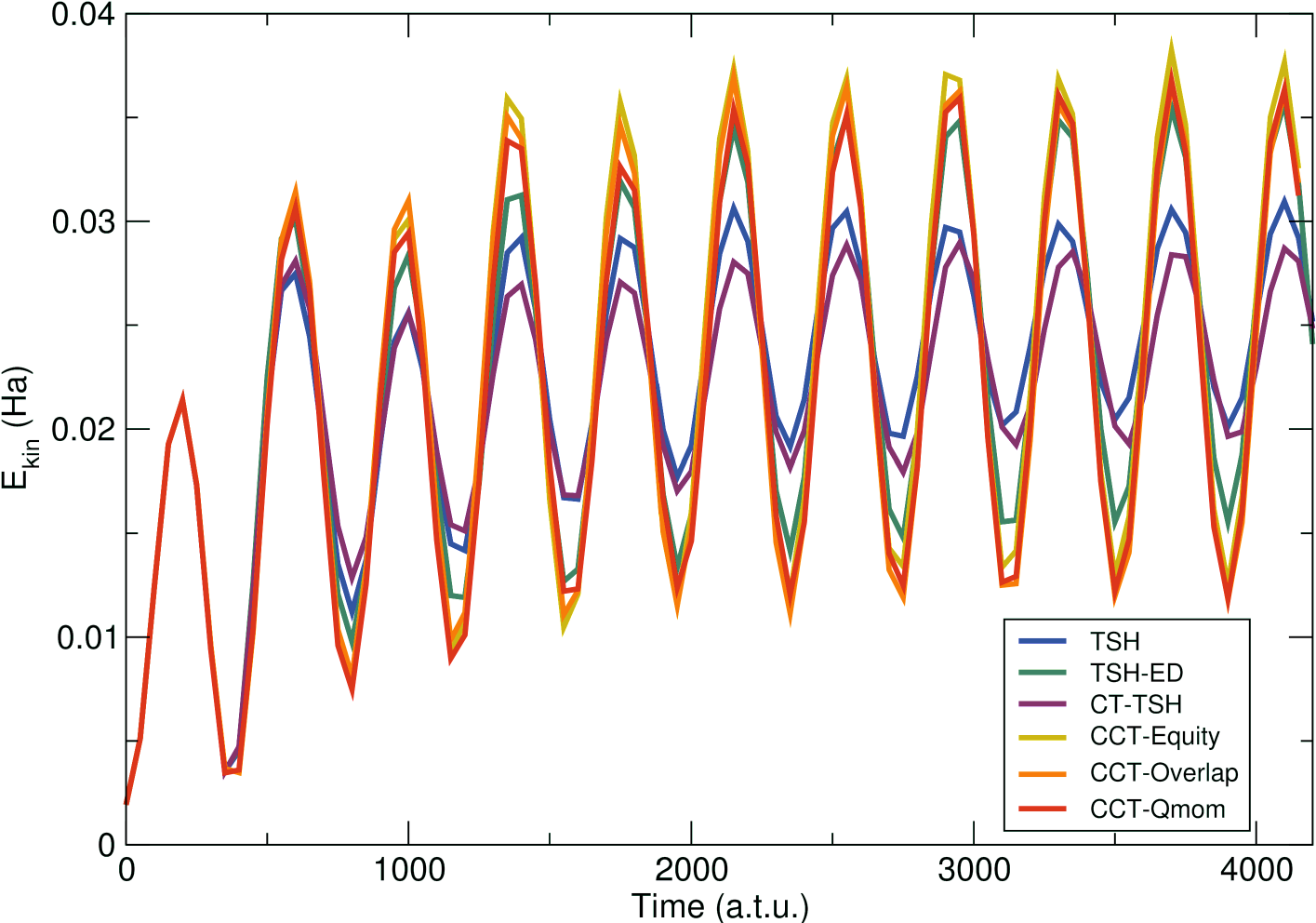}
\caption{Average kinetic energy of the ensemble of trajectories along the C=CH$_2$ stretching mode in fulvene.}
\label{fig:ekin}
\end{figure}

The kinetic energy along this normal mode (shown in Fig.~\ref{fig:ekin}) shows that the redistribution of kinetic energy in all flavors of CCT-TSH does not alter significantly the average kinetic energy along that normal mode. For TSH, TSH-ED and CT-TSH, the six different runs are shown, while for CCT-TSH we show, as in Fig.~\ref{fig:fulvene_pop}, the results obtained by reflecting the velocity of the trajectory that undergoes a frustrated hop (even though in the equity-based and overlap-based versions there are no frustrated hops) and by rescaling the velocity along the NACV.
The average kinetic energies of all methods are in good agreement and CCT-TSH yields values that lie within the uncertainty spread of the different standard TSH-ED schemes.

Having assessed the good performance and the stability of the energy-sharing schemes in fulvene on the excited-state population and on some aspects of the nuclear dynamics, we report additional comparisons of numerical schemes for CCT-TSH aiming to demonstrate its robustness.
\begin{figure}[hbt!]
\includegraphics[width=0.55\linewidth]{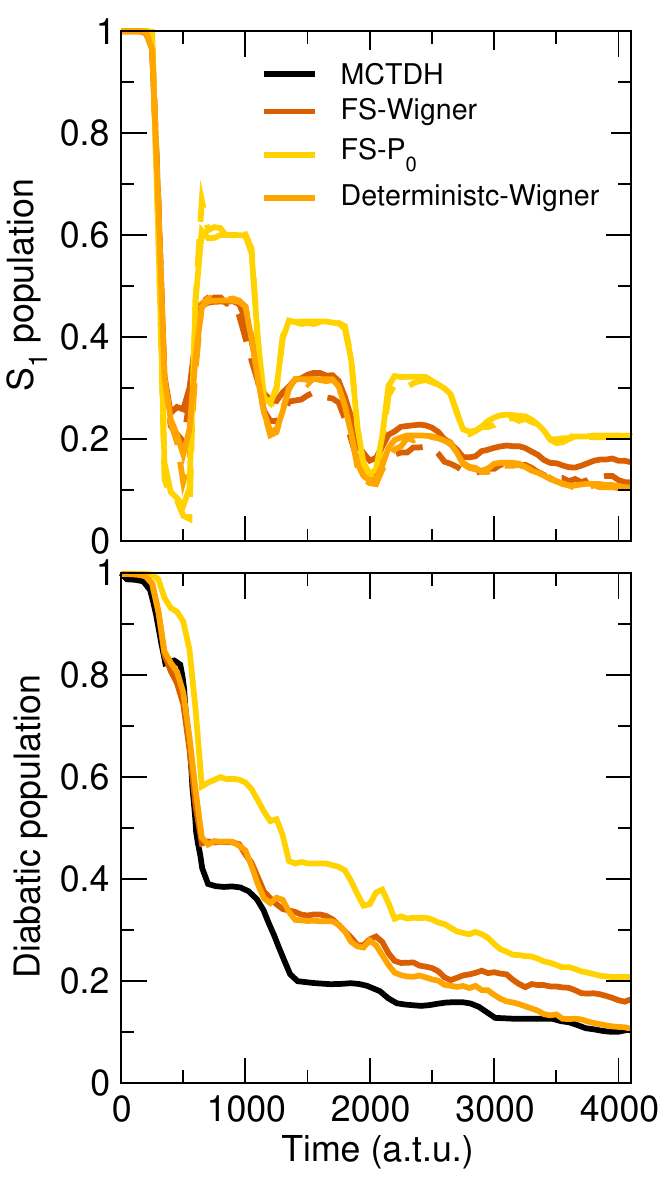}
\caption{Top panel: Population of S$_1$ in fulvene as a function of time calculated with CCT-TSH using different flavors of the overlap-based: using the fewest-switches procedure combined with a Wigner sampling of the initial conditions (in light-brown), using the fewest-switches procedure but sampling harmonically only the positions while setting all momenta of the trajectories at time $t=0$ to be equal to $\mathbf P_\nu=\mathbf P_0=\mathbf 0$ $\forall \nu$ (in yellow), using the deterministic selection of the active state combined with a Wigner sampling (in orange, as in Fig.~\ref{fig:fulvene_pop}). Solid lines show the average electronic populations (P) and dashed lines the fractions of trajectories (F). Bottom panel: Population of the diabatic state initially populated compared to MCTDH (in black) calculations of Ref.~[\!\!\citenum{gomez2024benchmarking}]}
\label{fig:more S1 pop fulvene}
\end{figure}
In Fig.~\ref{fig:more S1 pop fulvene}, we show the populations of S$_1$ (top panel) as function of time and we compare different flavors of the overlap-based energy sharing, namely by combining with the fewest-switches surface hopping (FS) to select the active state (in light-brown and in yellow in the figure) and either by using Wigner-sampled initial conditions (in light-brown and orange), i.e., positions and momenta, or by initializing the momenta to same zero value while sampling only the positions (in yellow). In Fig.~\ref{fig:more S1 pop fulvene}, we show as reference the results reported in Fig.~\ref{fig:fulvene_pop} where a deterministic selection of the active state is done by imposing the trajectories to follow the electronic state with largest population (in orange). While, as expected, the population slightly changes by changing the initial conditions, the internal consistency is not altered since no frustrated hops all along the simulated dynamics are observed (in all cases the velocity is rescaled along the NACV after the hop and there are no frustrated hops).

In Fig.~\ref{fig:more S1 pop fulvene} (bottom panel), we also report the population of the diabatic state that is fully populated at the initial time. We remind that CCT-TSH calculations are performed in the adiabatic basis, however, the LVC model is constructed in the diabatic basis. The QuantumModelLib library provides the transformation matrix from diabatic-to-adiabatic basis and vice versa, which depends on the nuclear positions. Therefore, at all nuclear positions, we are capable of transforming the electronic coefficients yielding the representation of the electronic conditional amplitude in the adiabatic basis to the diabatic representation. Nonetheless, this transformation provides a meaningful estimate of the diabatic population only if the algorithm is internally consistent. This is indeed the case for the CCT-TSH results shown in Fig.~\ref{fig:more S1 pop fulvene}. We compare the CCT-TSH diabatic populations to reference MCTDH results of Ref.~[\!\!\citenum{gomez2024benchmarking}]. The overlap-based scheme with deterministic selection of the active state shows an excellent agreement with the reference.

Another challenging test for the CCT-TSH algorithms is the S$_2$-to-S$_1$ photodynamics of DMABN. In this case, within the first 100~fs of dynamics, the system crosses multiple times nonadiabatic regions, such that the S$_2$ state loses population almost completely in a step-wise manner to S$_1$. We have recently shown~\cite{Agostini_JCP2025} that DMABN stretches the approximations underlying CT-MQC and CT-TSH, but CCT-TSH successfully cures the issues of the original coupled-trajectory algorithms guaranteeing internal consistency. Below, we explore the robustness of various implementations of CCT-TSH in this respect.

\begin{figure}[h]
\includegraphics[width=.6\linewidth]{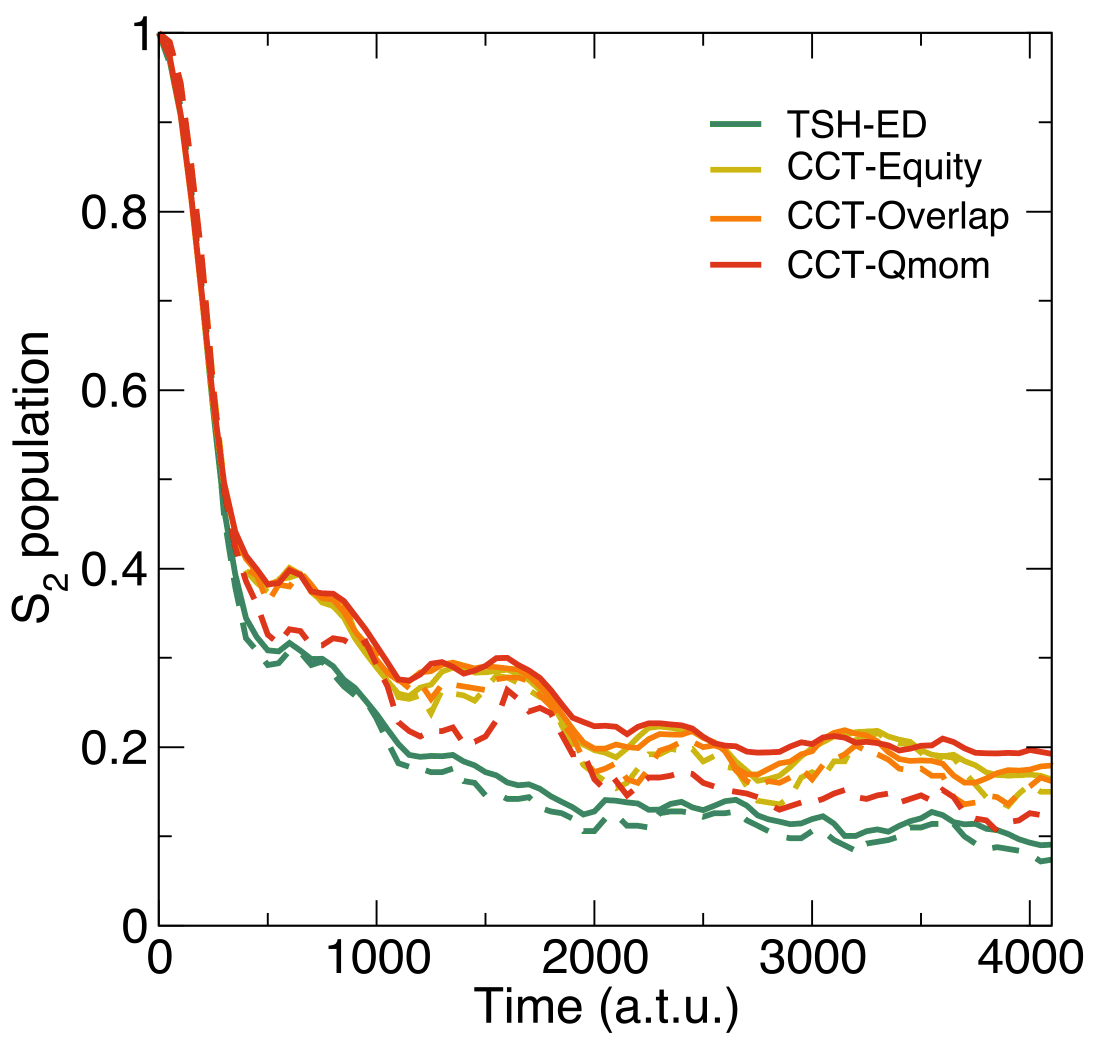}
\caption{Adiabatic population decays of S$_2$ in DMABN predicted by the dfferent surface hopping schemes, with TSH-ED (in green), with the equity-based scheme (in yellow), with the overlap-based scheme (in orange), with the quantum-momentum-based scheme (in red). Solid lines show the average electronic populations (P) and dashed lines
the fractions of trajectories (F).}
\label{fig:S1 DMABN}
\end{figure}

In Fig.~\ref{fig:S1 DMABN}, we report the population of the state S$_2$ as function of time, comparing CCT-TSH with TSH-ED. In all simulations, the velocity is rescaled along the NACV after the hop occurs and the velocity is inverted if a trajectory suffers from a frustrated hop. The population calculated using TSH-ED is shown in green in the figure, using dsahed lines for the fraction of trajectories (F) and continuous lines for the electronic populations (P). Note that in TSH-ED we observed a total of 529 frustrated hops. The equity-based (in yellow) and overlap-based (in orange) procedures for the energy sharing completely remove the occurrence of frustrated hops, yielding very similar populations, slightly larger than TSH-ED already after 500 a.t.u. The energy sharing with quantum-momentum-based scheme (in red) again produces 120 frustrated hops which does not seem to strongly affect the internal consistency. Nonetheless, the occurrence of frustrated hops suggests issues with the calculation of the quantum momentum, as observed previously for CT-MQC.~\cite{Agostini_JCP2025}

Finally, similarly to Fig.~\ref{fig:more S1 pop fulvene}, we compare additional flavors of the overlap-based energy-sharing scheme, namely in combination with fewest switches and with partial sampling of the initial conditions. We report the population of S$_2$ as function of time (top panel) and the population of the diabatic state that is almost completely populated at the initial time. We stress here that our trajectory-based calculations are initialized in the adiabatic state S$_2$, which has a mixed diabatic character in the Franck-Condon region. This feature explains the rapid population decrease of S$_2$ at very short times as well as the fact that the reported diabatic populations are not exactly one at the initial time. This difference in the initial condition might partly explain the deviation of the diabatic populations determined using CCT-TSH from the reference MCTDH results of Ref.~[\!\!\citenum{gomez2024benchmarking}]. Nevertheless, overall the agreement remains more than satisfactory.

\begin{figure}[hbt!]
\includegraphics[width=.55\linewidth]{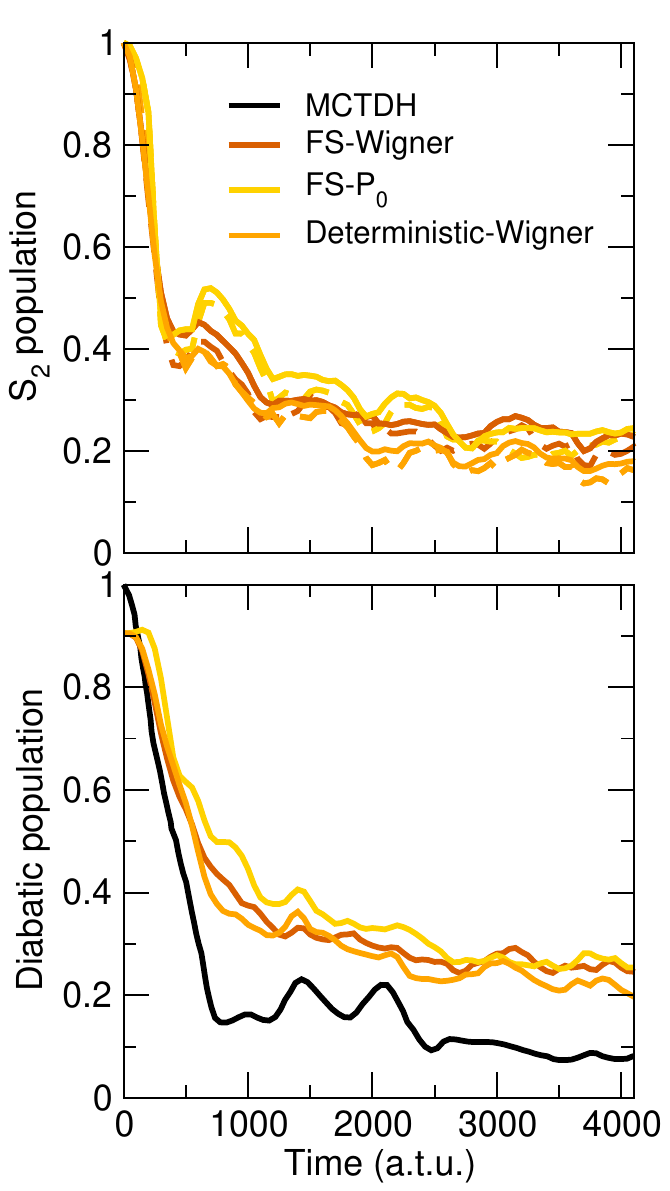}
\caption{Same as in Fig.~\ref{fig:more S1 pop fulvene} but for DMABN.}
\label{fig:more S1 pop DMABN}
\end{figure}

\section{Conclusions}\label{sec: conclusions}
In summary, we have identified a new and robust strategy to simulate the electron-nuclear dynamics underlying nonadiabatic processes, which is based on the exact factorization and employs the idea of surface hopping to approximate the nuclear dynamics. The crucial ingredient is the coupling of the trajectories, a concept that emerges naturally in the context of the exact factorization, and is the key to achieve internal consistency, to circumvent overcoherence, to alleviate the frustrated hops, and to produce stable results upon changing the velocity rescaling within surface hopping.

The coupling of the trajectories is essential to encode non-local, purely quantum mechanical features in the electronic and nuclear dynamics simulated using a surface hopping scheme, to calculate the quantum momentum and to introduce the energy sharing. Indeed, while the energy sharing is an idea grounded in the quantum-mechanical nature of the dynamics that the trajectories represent, its actual implementation remains somehow arbitrary. Three practical schemes were proposed and tested on full-dimensional photodynamics of fulvene and DMABN, based on linear vibronic coupling models. Out of these three flavors of CCT-TSH, two schemes, namely the equity-based and the overlap-based, were shown to give robust results for electronic and nuclear properties while reducing the occurences of frustrated hops, but we expect that alternative formulations might perform equally well.

The deterministic hopping strategy in CCT-TSH, guided directly by the time evolution of the electronic populations obtained from the exact factorization framework, proved to work quite well. By ensuring that nuclear trajectories evolve consistently in the electronic state with the largest population, the method maintains coherence between the electronic and nuclear dynamics. Our results demonstrate that this approach offers a robust and physically motivated alternative to the stochastic hopping schemes, with performance comparable to, or in some cases better than, the fewest-switches variant.


\begin{acknowledgement}
This work was supported by the French Agence Nationale de la Recherche via the projects Q-DeLight (Grant No. ANR-20-CE29-0014) and STROM (Grant No. ANR-23-ERCC-0002), and by a public grant from the Laboratoire d'Excellence Physics Atoms Light Matter (LabEx PALM) overseen by the French National Research Agency (ANR) as part of the ``Investissements d'Avenir'' program (reference: ANR-10-LABX-0039-PALM).
\end{acknowledgement}

\begin{suppinfo}
We provide as Supporting Information Python script (only the library tkinter is required to run the script) to help generate the input files to run all the calculations reported below, thus, all our data are easily reproducible. 
\end{suppinfo}


\providecommand{\latin}[1]{#1}
\makeatletter
\providecommand{\doi}
  {\begingroup\let\do\@makeother\dospecials
  \catcode`\{=1 \catcode`\}=2 \doi@aux}
\providecommand{\doi@aux}[1]{\endgroup\texttt{#1}}
\makeatother
\providecommand*\mcitethebibliography{\thebibliography}
\csname @ifundefined\endcsname{endmcitethebibliography}
  {\let\endmcitethebibliography\endthebibliography}{}

\end{document}